\documentclass[12pt]{article}

\begin{document}

\begin{titlepage}

\begin{flushright}
\end{flushright}
\vskip 2.5cm

\begin{center}
{\Large \bf Improved Bounds on Lorentz Symmetry Violation\\
From High-Energy Astrophysical Sources}
\end{center}

\vspace{1ex}

\begin{center}
{\large Brett Altschul\footnote{{\tt baltschu@physics.sc.edu}}}

\vspace{5mm}
{\sl Department of Physics and Astronomy} \\
{\sl University of South Carolina} \\
{\sl Columbia, SC 29208} \\
\end{center}

\vspace{2.5ex}

\medskip

\centerline {\bf Abstract}

\bigskip

Observations of the synchrotron and inverse Compton
emissions from ultrarelativistic electrons in astrophysical sources
can reveal a great deal about the energy-momentum relations of those electrons. They can thus
be used to place bounds on the possibility of Lorentz violation in the electron sector.
Recent $\gamma$-ray telescope data allow the Lorentz-violating electron
$c^{\nu\mu}$ parameters to be constrained extremely well,
so that all bounds are at the level of $7\times 10^{-16}$ or better.

\bigskip

\end{titlepage}

\newpage

\section{Introduction}

Since Einstein's first 1905 publications, the special theory of relativity has proven itself to be an
extremely accurate and durable framework; in conjunction with quantum field theory, it provides an
apparently complete picture of all non-gravitational physics. Moreover, the Lorentz invariance of the
special theory also undergirds general relativity (GR) as well, where it plays the role of a
local symmetry.

However,
in the last quarter century, there has been a real renewal of interest in the possibility that
Lorentz symmetry might not be absolute, but only approximate. While there have been sporadic reports
of marginal evidence for Lorentz violation, there has thus far been no lasting and compelling experimental
evidence that Lorentz symmetry is not absolute. However, it is still worthwhile to have a comprehensive
understanding of how well Lorentz invariance---like any important symmetry---has been tested. Much of the
story of modern foundational physics has been about how transformations that may initially have appeared to
be exact symmetries of nature were found to be ever so slightly violated; for example: isospin symmetry, parity
(P), charge conjugation (C), and time reversal (T).
The discoveries of the ways in which each of these symmetries is broken
led to important new insights into the structure of fundamental physics.
If violations of some components of Lorentz symmetry (meaning
spatial isotropy and Lorentz boost invariance) were
similarly uncovered, that would be another critically important discovery. It would change our understanding
of how the laws of physics operate at the deepest levels; moreover, the discovery would automatically provide a
new experimental method for studying these truly novel realms of physics. Conversely, confirmations that Lorentz
symmetry remains valid even in extremely precise measurements can give us insight into the forms that new
fundamental physics cannot take without being inconsistent with experiment. These are all important reasons to
have a program of strong and systematic measurements of Lorentz invariance.

Moreover, there are other reasons to think that studying the possibility of Lorentz violation may be fruitful.
Probably the deepest remaining puzzle in fundamental physics is the quantization of gravity. Naive methods of
studying quantum gravity fail because the theory is not renormalizable, and so some kind of alternative approach
is needed. There have been numerous attempts to develop schematic theories of quantum gravity, and, quite
interestingly, many of these speculative frameworks appear to include Lorentz-violating regimes---although
whether those regimes are physically plausible cannot be answered with our current levels of understanding. So
it may turn out that investigations of Lorentz violation will be a window into understanding quantum gravity.

Lorentz symmetry is also closely tied to another symmetry that is---so far as we know---exact in nature.
This is the combined action of the discrete symmetries C, P, and T. Even with possibly non-local
interactions, CPT violation in an interpretable quantum field theory (QFT) would automatically entail Lorentz
violation~\cite{ref-greenberg}; the only additional technical requirement connecting Lorentz and CPT symmetries
is the existence of a well-defined $S$-matrix. (Note, however, that these arguments do not apply to a geometric
gravity theory like GR, which is not a QFT defined in terms of an $S$-matrix.) This means that many tests of
Lorentz invariance may be reinterpreted as CPT tests, and vice-versa. However, it is important to emphasize
that the 
connection between the two types of symmetries is not symmetric. It is perfectly possible to have Lorentz
violation without affecting CPT
symmetry, and indeed, in this work  we shall concentrate on some specific forms
of Lorentz violation that do not also break CPT symmetry.

The modern systematic study of Lorentz (and CPT) symmetry violations was made possible by advances in
effective field theory (EFT) methods in recent decades. With the machinery of EFT, it became possible to
write down test theories that can accommodate deviations from these symmetries, in the context of either
standard model particle physics or gravitational phenomena. The generic effective field theory that can
describe the symmetry-breaking effects is known as the standard model extension (SME).
The operators in the SME action resemble those that appear in the usual standard model, the key
difference being except that the field operators may have residual Lorentz tensor indices, which are
contracted not
with other field variables but with externally prescribed constants~\cite{ref-kost1,ref-kost2}.
These prescribed constants indicate
preferred vectors and tensors that exist in the vacuum; the universe can be endowed with preferred spacetime
directions, even when there is no matter present. If the ultimate origin of Lorentz violation is through
spontaneous symmetry breaking (the same way that the gauge symmetry of the standard model electroweak sector
is broken), then the tensor-valued coefficients will be proportional to the vacuum expectation values of
tensor-valued
dynamical fields.
Because it is so general, the full SME necessarily must contain an infinite number of Lorentz-violating
operators in its action; however, in the particle physics sector, these operators may be ordered by mass
dimension in the same was as the Lorentz-invariant operators of the standard model. The analogue of the
renormalizable standard model---containing only
local operators constructed out of standard model fields, which
are additionally gauge invariant, translation invariant, and superficially power-counting renormalizable---is
the minimal SME. Since it contains only a finite number of novel coupling constants, the minimal SME makes
a natural test theory framework for analyzing many experimental tests of Lorentz invariance. 
Experiments  drawn from many
disparate areas of physics can be used to place bounds on the various coefficients in the minimal SME, and
the best current bounds have been systematically tabulated in~\cite{ref-tables}.

Astronomical measurements can be useful sources of information about Lorentz symmetry, and many of the best
bounds on SME parameters have historically been derived from analyses of astronomical data. The main advantages
of looking at data from outer space are twofold; experimental sensitivities may be enhanced by
the presence of extremely long lines of sight or extremely high particle energies.
Thanks to these factors, photons coming to us from
distant astrophysical sources can become powerful probes of new physics.
By looking at the spectrum of photons emitted by faraway sources, we may learn a
lot about the energy-momentum of the photons themselves. Deviations from the expected form
$E_{\gamma}=|\vec{p}\,|$ (especially if the modified dispersion relations are polarization dependent) would
often have clear signatures. However, photon observations can tell us about more than just the behavior of
freely-streaming photons. The facts that a photon was emitted at all, and that it traveled along a certain
direction to reach our detector, are also important, and these facts can reveal additional information about
other particle species that the photon may have interacted with. Of particular importance in the present
analysis will be production and propagation of extremely energetic (TeV-scale) $\gamma$-rays.

The highest energy $\gamma$-rays observed from energetic astrophysical sources are typically produced through
one of two processes, involving even-more-energetic leptons or hadrons. The leptonic process is
inverse Compton (IC) scattering---the upscattering of a comparatively low-energy photon through a collision
with an ultrarelativistic electron, $e^{-}+\gamma\rightarrow e^{-}+\gamma$. The
main hadronic process
for producing the $\gamma$-rays is neutral pion decay, $\pi^{0}\rightarrow\gamma+\gamma$, where the neutral
pions are typically themselves produced through prior collisions involving longer-lived baryons. The very
fact that such $\gamma$-rays can be produced at all actually reveals quite a bit about the energy-momentum
relations for the parent particles in these reactions, as well as the daughter photons.
Experimental observations of the spectra of astrophysical
$\gamma$-ray sources may thus be used to place bounds on Lorentz violation in both the lepton and hadron
sectors of the SME~\cite{ref-jacobson1,ref-altschul6,ref-altschul15,ref-altschul14,ref-altschul16}, although our
principal focus here will be on the leptonic sector.

Unsurprisingly, the best SME bounds are attained
when a source's structure and emission spectrum are well understood, so that we may draw reliable inferences
about what processes are responsible for the production of the bulk of the source's emissions; and yet, even
when the production mechanisms for the TeV $\gamma$-rays are unclear, there are other inferences we may draw,
just based on the observed fact that those photons have traversed interstellar space and reached our
detectors. Lorentz violation opens up the possibility that there may be
photon decay or disappearance processes that would have been absolutely forbidden in a Lorentz-invariant
theory---such as $\gamma\rightarrow e^{+}+e^{-}$ in vacuum. When there is no other particle nearby to catalyze
the process by absorbing some of the photon's momentum, this process is impossible in the conventional standard
model, since the initial energy $E_{\gamma}=|\vec{p}\,|$ will never be sufficient to create two massive daughter
particles with the same total momentum $\vec{p}$. However, if the ultrarelativistic dispersion relations for
the electron and positron (or, in fact, for any charged particle-antiparticle pair) grow more slowly as
a function of momentum than in the usual Lorentz invariant case, then the decay process may become allowed above
a certain photon threshold energy. (There is also a possible hadronic process by which TeV $\gamma$-rays may
lose energy above a certain threshold if there is Lorentz violation in the hadron sector:
$\gamma\rightarrow\gamma+\pi^{0}$. However, we shall not consider this hadronic process in the present
analysis. Nor shall we be concerned with the photon splitting process $\gamma\rightarrow N\gamma$, which
is normally forbidden through a different mechanism, related to gauge invariance
of the necessary on-shell matrix element~\cite{ref-schwinger1}. Radiative corrections in the SME may
make this process possible, even without any explicit Lorentz violation in the photon sector~\cite{ref-kost5}.
However, in this case the splitting process is automatically exactly at threshold, which leads to phase
space issues, and so the process would not manifest itself as a single decay but rather as a coherent
oscillations process, along the lines of $\gamma\rightarrow N\gamma\rightarrow\gamma\rightarrow\cdots$.)

The remainder of this paper will be organized in the following fashion. In section~\ref{sec-SME}, we shall
introduce the minimal SME terms that modify the energy-momentum relations for ultrarelativistic fermions
and examine how the modified particle kinematics lead to changes in the fermions' interactions with
the electromagnetic field. Section~\ref{sec-bounds} turns this around to show how observations of x-rays and
$\gamma$-rays from multiple sources can be used to place strong bounds on certain SME coefficients. Finally,
section~\ref{sec-concl} summarizes our main conclusions and discusses possibilities for further work in this
area.

\section{Relativistic Particle Energies in the Minimal SME}
\label{sec-SME}

The Lagrange density for the electron-positron sector of the minimal SME is
\begin{equation}
\label{eq-L}
{\cal L}=\bar{\psi}(i\Gamma^{\mu}\partial_{\mu}-M)\psi.
\end{equation}
The term $M$ contains not just the scalar mass, but Dirac matrices of all the possible tensor types.
However, the Lorentz violation coefficients that are collected in $M$ all resemble the mass, in that they
become relatively unimportant contributors to energy-momentum relation for electrons or positrons at
ultrarelativistic energies. In contrast, the two minimal SME tensors that appear in $\Gamma^{\mu}$
govern the Lorentz-violating effects that predominate at those energies. These terms are
\begin{equation}
\label{eq-Gamma}
\Gamma^{\mu}=\gamma^{\mu}+c^{\nu\mu}\gamma_{\nu}+d^{\nu\mu}\gamma_{5}
\gamma_{\nu}.
\end{equation}
Just looking at the Lorentz structure of $\Gamma^{\mu}$, it may be observed that additional Dirac
matrices could also be included on the right-hand side of (\ref{eq-Gamma}).
For general fermions, this is true, and there are additional vector,
axial vector, and three-index tensor terms to be included. However, for real electrons, such terms are
not part of the minimal SME, because they violate $SU(2)_{L}$ gauge invariance by mixing right-handed and
left-handed fields.

The effects of the pair of traceless two-index tensors in (\ref{eq-Gamma}) on particles with very large energies
can most easily be expressed in terms of the maximum achievable velocities (MAV) for the particles. (As
indicated above, the neglected terms that comprise $M$ do not affect particle MAVs.) The MAV for a particle
depends, in general, on the direction of motion, the spin helicity, and the whether the particles is
a fermion or antifermion. The maximum speed in a direction $\hat{p}$ is given by
\begin{equation}
\label{eq-MAV}
v_{{\rm MAV}}(\hat{p})=
1+\delta(\hat{p})=1-c_{00}-c_{(0j)}\hat{p}_{j}-c_{jk}\hat{p}_{j}\hat{p}_{k}+sd_{00}+sd_{(0j)}
\hat{p}_{j}+sd_{jk}\hat{p}_{j}\hat{p}_{k};
\end{equation}
here, the parentheses indicate a symmetrization with respect to the two indices, as
in $c_{(0j)}=c_{0j}+c_{j0}$, and $s$ encapsulates the dependences on the spin and the particle type.
The quantity $s$ is the product
of the particle helicity and its fermion number ($+1$ for fermions like electrons, versus $-1$ for
antifermions). In (\ref{eq-MAV}) and in what follows, we are neglecting all terms of higher than
first order in the Lorentz violation coefficients, because Lorentz violation, if it exists physically, is
established to be a small effect.

The MAV (\ref{eq-MAV}) is a consequence of an ultrarelativistic dispersion relation
\begin{equation}
\label{eq-EofP}
E=\sqrt{m^{2}+p^{2}[1+2\delta(\hat{p})]},
\end{equation}
although this expression also neglects some terms that are suppressed ultrarelativistically (that is,
suppressed by higher powers of $m/p$). The kinematics of high-energy electrodynamic
processes are determined by this dispersion relation,
in conjunction with the dispersion relations for the photon field.
However, there are also other effects we shall be interested in, whose behavior is not just governed
by threshold kinematics. To understand these phenomena, we need to know a bit more about the velocity,
which takes the form, at ultrarelativistic energies, of
\begin{equation}
\vec{v}=\left[1+\delta(\hat{p})-\frac{m^{2}}{2E^{2}}\right]\hat{p}=\vec{v}_{{\rm MAV}}(\hat{p})
-\frac{m^{2}}{2E^{2}}\hat{p}.
\end{equation}
Note also that this expression is simply the usual ultrarelativistic velocity plus
$\delta(\hat{p}\,)\hat{p}$.

At noted above, to understand the kinematics of electromagnetic interactions at extreme energies,
knowledge of the photon dispersion relations are also needed. The key expressions for the photons
energies are actual rather similar to the corresponding fermionic expressions, although there is a somewhat
richer spin structure. On the other hand, the photon dispersion relations are also simplified somewhat
by the fact that the quanta, as they are massless, will always move at their MAV.
Moreover, the spin dependence of the photon MAV is actually extraordinarily well constrained experimentally,
because a spin-dependent electromagnetic phase speed means vacuum birefringence---something which has
been searched for and not seen, even for photons that have traveled cosmological
distances~\cite{ref-kost21,ref-mewes5}.

The remaining non-birefringent terms in the photon MAV look just like the $s$-in\-de\-pen\-dent part of
(\ref{eq-MAV}). However, we can also avoid dealing with these terms explicitly, through a clever choice of
coordinates. It is always possible, through a linear transformation of coordinates, to remove the
spin-independent, dimension-four Lorentz-violating terms from the propagation Lagrangian for one standard
model field, at the cost of creating analogous terms in the Lagrangians for all other species. In many
cases, including in this paper, it is most convenient to make the Maxwell sector conventional.
The coordinate freedom (combined with the observed absence
of vacuum birefringence) allows us to set $\delta_{\gamma}=0$.
This means that we take the freely-propagating electromagnetic field as representing the ``rods and clocks''
that are used to define our measurement coordinates, which simplifies some calculations quite a bit.
Moreover, bounds based purely on high-energy interactions between the electron field and the photon field are
actually easy to transform between the different coordinate systems. The constraints on the electron
$c^{\nu\mu}$ that we shall derived based on ultrarelativistic reaction thresholds translate
into (in coordinates
for which the electromagnetic sector is not automatically assumed to be conventional) bounds on differences
$c^{\nu\mu}-\frac{1}{2}(k_{F})_{\alpha}\,^{\mu\alpha\nu}$ of electron and certain
specific photon SME coefficients.

Some phenomenalistic consequences of having a charged particle MAV that need not be the same as the
phase speed of light are obvious. For instance, if an electron's speed is greater than 1, then there will
have to be vacuum Cerenkov emission. (The vacuum Cerenkov process behaves very much like regular Cerenkov
emission in matter---where it originates from charged particles exceeding the speed of the light in the
material medium.) This simple observation is actually the key to many of the bounds that we shall be able to
set.

However, some other consequences of the revised energy-momentum relations are more subtle. In particular,
understanding the synchrotron radiation spectrum from the highest-energy photons requires understanding
of the velocity beyond just the MAV. One of the consequences of gauge invariance is that the electromagnetic
field always couples to charged currents
solely through the velocity; this means that if the trajectory of a moving charge is known, then the
radiation it emits is also completely known, regardless of whatever modified dynamics may
have been responsible for
determining the charge's motion in the first place. The radiation is also beamed into a very narrow pencil of
angles around the instantaneous direction that the charge is moving. An immediate consequence of the
beaming is that
whatever radiation we observe from a source must have been emitted by electrons that were moving in the
source-to-Earth direction.  If the synchrotron spectrum of a particular source
indicates the presence of a population of electrons with Lorentz factors up to $\gamma_{\max}$
[and by the Lorentz factor $\gamma$, we mean precisely $\gamma=(1-v^{2})^{-1/2}$,
a quantity which depends only on a particle's velocity, not directly on its energy or momentum],
then there must be electrons moving toward us with speeds at least 
\begin{equation}
v=1-\frac{1}{2\gamma_{\max}^{2}}
\end{equation}
Consequently, the MAV for those electrons must be at least this large, or
\begin{equation}
\label{eq-synclimit}
c_{00}+c_{(0j)}\hat{e}_{j}+c_{jk}\hat{e}_{j}\hat{e}_{k}
-\left|d_{00}+d_{(0j)}\hat{e}_{j}+d_{jk}\hat{e}_{j}\hat{e}_{k}\right|<\frac{1}{2\gamma_{\max}^{2}},
\end{equation}
where the unit vector $\hat{e}$ is pointing in the source-to-Earth direction.

There are two things that make applying the bound (\ref{eq-synclimit}) somewhat tricky. First, the
bound in strictly one sided; it only limits the possibility that the MAV may be less than~1. Second,
there only needs to be one helicity of electrons for which the Lorentz factor $\gamma_{\max}$ is reached.
It is because the $d^{\nu\mu}$ terms always speed up one spin state and slow down the other that they appear in
(\ref{eq-synclimit}) in the undesirable
fashion they do. Without independent bounds on the $d^{\nu\mu}$, it is not possible
to extract bounds on the $c^{\nu\mu}$ from (\ref{eq-synclimit}). Fortunately, most of the $d^{\nu\mu}$
coefficients are accessible in other kinds of experiments. In particular, the $d^{\nu\mu}$ coefficients
governing
anisotropy and anisotropic boost violation---that is, all the components of the tensor except for
$d^{00}$---have been measured using torsion pendulum
experiments with test masses carrying macroscopic spin polarizations~\cite{ref-heckel3}.
These are never ``clean'' bounds, in that
they involve not just the $d^{\nu\mu}$, but also linear combinations of other SME coefficients. However, the
constraint scales they imply for the various $d^{\nu\mu}$ are all at the $10^{-26}$ level or better; these
bounds are thus several orders of magnitude tighter than any that we shall be able to place astrophysically.
It is thus reasonable to neglect the $d^{\nu\mu}$ coefficients, just as we have similarly neglected any
spin-dependent terms in the photon sector, because the are already so tightly bounded. The obvious potential
exception is with the pure isotropic boost symmetry violation term $d^{00}$, for which there has not been
an analysis placing rigorous bounds. However, the torsion pendulum bounds on other forms of boost invariance
violation are so much tighter than even the best high-energy astrophysical bounds, that any electron
$d^{00}$ that was large enough to have an observable affect on synchrotron (or IC) spectra would
presumably have led to large anomalies in the torsion pendulum data, which ought to have easily been noticed.
So it is reasonable step to neglect even the $d^{00}$ term in (\ref{eq-synclimit}) and elsewhere---although it
would certainly also be desirable to see an analysis of the torsion measurement experiments that did
explicitly bound $d^{00}$.

Complementary to the one-sided synchrotron bounds derived from (\ref{eq-synclimit}) are bounds coming
from the observed presence of IC $\gamma$-rays coming from various sources. For the IC process
$e^{-}+\gamma\rightarrow e^{-}+\gamma$ to have time
to occur on a significant scale, the ultrarelativistic electrons must live long enough to collide with
low-energy photons (which may be from the optical, infrared, or radio spectra; synchrotron radiation itself
provides an important pool of photons to be IC upscattered). If the electrons are moving faster than 1 (the
phase speed of light), then they will lose energy through vacuum Cerenkov emission---which is an even more
efficient radiation process than IC scattering. Thus, if the energy-momentum relation (\ref{eq-EofP})
has $\delta(\hat{p})>0$, allowing for sufficiently energetic electrons to be superluminal, there will be a
natural cutoff in the IC spectrum, since electrons above a certain $E$ will not keep their energy long
enough to undergo the IC interaction process. Again neglecting the effects of $d^{\mu\nu}$, the condition
for electrons up to an energy $E_{\max}$ to be subluminal when moving in the $\hat{e}$-direction is
\begin{equation}
\label{eq-IClimit}
c_{00}+c_{(0j)}\hat{e}_{j}+c_{jk}\hat{e}_{j}\hat{e}_{k}>-\frac{1}{2(E_{\max}/m)^{2}}.
\end{equation}
Obviously, these limits (which exclude MAV values that are greater than 1) are perfectly complementary
to the synchrotron spectrum bounds.

Moreover, there is also one more way that the observation of IC $\gamma$-rays can constrain a MAV smaller
than 1. In fact, as previously, noted, the survival of any $\gamma$-ray, whether it was produced through the
leptonic IC process or through a hadronic decays, tells us something about the energy-momentum relations
for electrons and positrons. The absence of the QED process $\gamma\rightarrow e^{+}+e^{-}$ (which would
be exceedingly rapid if it were allowed) for a photon with energy $E_{\gamma}$ traveling in the direction
$\hat{e}$ implies that
\begin{equation}
\label{eq-decaylimit}
c_{00}+c_{(0j)}\hat{e}_{j}+c_{jk}\hat{e}_{j}\hat{e}_{k}<\frac{2}{(E_{\gamma}/m)^{2}}
\end{equation}
Moreover, while it is slightly less sensitive, the limit (\ref{eq-decaylimit})
is genuinely independent of the $d^{\nu\mu}$ coefficient and does not rely on external
constraints on the $d^{\nu\mu}$. The reason is that, when the quanta in the reaction are all collinear (as they
are at threshold), the electron and positron must have the same helicity, since the parent photon was a spin-1
state. The $d^{\nu\mu}$ terms then shift the energies of the electron and positron daughters in opposite
directions, making the overall threshold location independent of $d^{\nu\mu}$.

\section{Empirical Bounds on SME Parameters}
\label{sec-bounds}

From what we see of the emissions from energetic astrophysical sources, we can infer quite a bit of useful
information about Lorentz symmetry. The absence of anomalous features in a source's spectra may be
used to place strong bounds on the SME coefficients—and, as we have seen, this is particular the case for the
coefficients that compose the electron $c^{\nu\mu}$ tensor. However, even for the components of $c^{\nu\mu}$,
there are some terms that are much better constrained by laboratory experiments.

For comparing bounds on the SME coefficients, it is necessary to have a single agreed-upon laboratory
coordinate system in which the bounds are to be expressed. This is conventionally a Sun-centered
celestial equatorial coordinate system~\cite{ref-bluhm4}.
The coordinate origin lies at the center of the Sun, and the $Z$-direction lies parallel to the Earth's axis.
The $X$-axis points toward the location of the vernal equinox on the celestial sphere, and the
direction of the $Y$-axis is then determined the right hand rule. The temporal coordinate is $T$, with its epoch
at the beginning of the year 2000 (although the zero point rarely matters).

The best current laboratory bounds on the electron $c^{\nu\mu}$ come from measurements with a Yb$^{+}$
optical-frequency clock~\cite{ref-sanner}. The bounds on four of the coefficients that describe anisotropy in
a stationary lab frame ($c_{XY}$, $c_{XZ}$, $c_{YZ}$, and $c_{-}=c_{XX}-c_{YY}$) have been bounded
both above and below at the $10^{-20}$ level. These bounds are not quite as clean as astrophysical threshold
bounds, since the introduction of the clock timing apparatus gets the experimental observables somewhat
entangled with
the effective $c^{\nu\mu}$ coefficients for ordinary bulk matter. However, these bounds are still strong enough
that we may reasonably set these four particular coefficients to zero in our analyses. There are also
Yb$^{+}$ clock bounds on the boost anisotropy coefficients $c_{TX}$, $c_{TY}$, and $c_{TZ}$, although
these are less sensitive, because they rely on the relatively slow orbital velocity of the Earth to test
different reference frames. (In contrast, one of the key advantages of the ultrarelativistic emission bounds
is their roughly equal sensitivities to all nine of the $c^{\nu\mu}$ components.) There are also few good
bounds on the fifth stationary anisotropy parameter, $c_{Q}=c_{XX}+c_{YY}-2c_{ZZ}$, since $c_{Q}$ does
not generate any energy shifts that change as the Earth rotates daily on its axis.

The astrophysical observations will therefore be most useful for placing bounds on $c_{TT}$,
$c_{TX}$, $c_{TY}$, $c_{TZ}$, and $c_{Q}$.
When only these five coefficients are nonzero, The MAV modification which appears in the bounds
(\ref{eq-synclimit}--\ref{eq-decaylimit}) becomes
\begin{equation}
\label{eq-reduceddelta}
-\delta(\hat{e})=\frac{4}{3}c_{TT}+2c_{TX}\hat{e}_{X}+2c_{TY}\hat{e}_{Y}+2c_{TZ}\hat{e}_{Z}
-\frac{1}{2}c_{Q}\left(\hat{e}_{Z}^{2}+\frac{1}{3}\right).
\end{equation}
(To get this expression, we have used the symmetry $c^{\nu\mu}=c^{\mu\nu}$ and tracelessness
$c^{\mu}\,_{\mu}=0$ of the tensor.)
Obviously, in order to place rigorous two-sided bounds on the five individual $c^{\nu\mu}$ parameters in
(\ref{eq-reduceddelta}), it is necessary to have data from multiple sources; at least six
inequalities, coming from observations of at least five different sources are required, in fact.
With a sufficient number of inequalities, the coefficients will be constrained to lie
in a bounded region of parameter space, and
the bounds on individual coefficients can then be extracted by linear programming.

However, it has also become somewhat common to express bounds in terms of the ``maximum 
reach'' of an experimental method with respect to each particular SME parameter. This is done by
working in the context of an artificial test model in which only a single SME coefficient is allowed
to be nonzero; if our case, that consists of setting four of the five coefficients
$c_{TT}$, $c_{TX}$, $c_{TY}$, $c_{TZ}$, and $c_{Q}$ to zero and then looking at the experimentally allowed
values of the remaining nonzero quantity.

\begin{table}
\begin{center}
\begin{tabular}{|l|c|c|c|c|c|}
\hline
Emission source & $\hat{e}_{X}$ & $\hat{e}_{Y}$ & $\hat{e}_{Z}$ &
$\gamma_{\max}$ & $E_{\max}/m$ \\
%
%
\hline
Centaurus A & 0.68 & 0.27 & 0.68 & $2\times 10^{8}$\cite{ref-kataoka} & - \\
Crab nebula & $-0.10$ & $-0.92$ & $-0.37$ & $3\times
10^{9}$\cite{ref-aharonian1} & $7\times 10^{8}$~\cite{ref-amenomori} \\
eHWC J$1825 - 134$ & $-0.11$ & 0.97 & 0.23
& - & $2\times 10^{8}$\cite{ref-abeysekara,ref-abdalla} \\
eHWC J$1907 + 063$ & $-0.29$ & 0.95 & $-0.11$
& - & $2\times 10^{8}$\cite{ref-abeysekara} \\
eHWC J$2019 + 368$ & $-0.46$ & 0.66 & $-0.60$ 
& - & $2\times 10^{8}$\cite{ref-abeysekara} \\
MSH 15-52 & 0.34 & 0.38 & 0.86 &
- & $8\times 10^{7}$\cite{ref-aharonian4} \\
RCW 86 & 0.35 & 0.30 & 0.89 &
$10^{8}$\cite{ref-rho} & - \\
Vela SNR & 0.44 & $-0.55$ & 0.71 & $3\times 10^{8}$\cite{ref-mangano} &
$1.3\times 10^{8}$\cite{ref-aharonian2} \\
\hline
\end{tabular}
\caption{
\label{table-sources}
Location and spectrum parameters for the sources used to constrain the $c^{\nu\mu}$.
References are given for each value of
$\gamma_{\max}$ or $E_{\max}$.}
\end{center}
\end{table}

For placing two-sided bounds, there are two key pieces of potentially useful information about a source.
One is the highest energies of the IC $\gamma$-rays seen coming from the source. The $\gamma$-ray energy
immediately provides a survival bound of the form (\ref{eq-decaylimit}). Moreover, a known IC $\gamma$-ray
also provides another bound, derived from (\ref{eq-IClimit}), when we make use of the fact that
the observed $\gamma$-ray cannot have more energy than the original ultrarelativistic lepton that
upscattered it, or $E_{\max}>E_{\gamma}$. The other potentially important data point for each source is
the value of $\gamma_{\max}$ that we may infer separately from the synchrotron portion of the
source's photon spectrum.
Sources' $\gamma_{\max}$ values are a bit more subtle than the simple energy values for the observed
IC photons. Each $\gamma_{\max}$ must be extracted from a model of the source, and this makes it
critical that only thoroughly well-understood sources be used.
The most significant feature in the upper range of a typical synchrotron spectrum is a frequency cutoff
$\nu_{c}$. Above $\nu_{c}$, the emission falls off exponentially, and this makes it relatively
straightforward to derive a lower limit on $\nu_{c}$ from an observed spectrum. From $\nu_{c}$,
it is then possible to infer $\gamma_{\max}\propto\sqrt{\nu_{c}/B_{\perp}}$, with $B_{\perp}$ being
the component of the magnetic field at the source that lies perpendicular to the line of sight. The
field $B_{\perp}$ may be estimated fairly easily based on the low-frequency part of the
synchrotron spectrum in combination with simple energetic consideration, and the overall procedure for placing
a lower limit on $\gamma_{\max}$ is quite robust. The inferred values of $E_{\max}$ and $\gamma_{\max}$
for a number of useful and well-studied sources are given in table~\ref{table-sources}. Information
about the IC origins of some of the most recently observed $\gamma$-rays may be found
in~\cite{ref-fang,ref-dimauro}.

\begin{table}
\begin{center}
\begin{tabular}{|c|c|c|c|c|}
\hline
& Coupled & Coupled & Reach & Reach \\
$c^{\mu\nu}$ & Maximum & Minimum & Maximum & Minimum \\
\hline
$c_{TT}$ & $7\times 10^{-17}$ & $-1.1\times 10^{-16}$ & $4\times 10^{-20}$ & $-8\times10^{-19}$ \\
$c_{TX}$ & $6\times 10^{-17}$ & $-7\times 10^{-16}$ & $5\times 10^{-18}$ & $-3\times10^{-19}$ \\
$c_{TY}$ & $2\times 10^{-17}$ & $-1.1\times 10^{-16}$ & $6\times 10^{-19}$ & $-3\times10^{-19}$ \\
$c_{TZ}$ & $4\times 10^{-16}$ & $-5\times 10^{-17}$ & $1.4\times 10^{-18}$ & $-8\times10^{-20}$ \\
$c_{Q}$ & $3\times 10^{-16}$ & $-3\times 10^{-16}$ & $4\times 10^{-18}$ & $-3\times10^{-19}$ \\
\hline
\end{tabular}
\caption{
\label{table-bounds}
Astrophysical bounds on the components of $c$. The second and third columns give the maximum and minimum
allowed values that are possible when all five the $c^{\nu\mu}$ coefficients in the table are permitted
to be nonvanishing. The fourth and fifth columns give the maximum experimental reach of the astrophysical
data---the bounds that result when only one coefficient at a time is allowed to be nonzero.}
\end{center}
\end{table}

The resulting bounds on the five relevant $c^{\nu\mu}$ parameters are listed in table~\ref{table-bounds}.
There are a number of improvements over previous astrophysical measurements. These are based on
ongoing improvements to the TeV $\gamma$-ray spectra coming from the most recent generations of
telescopes. However, some of the improvement is also due indirectly to the influence of the terrestrial
atomic clock experiments, which we used to narrow the parameter space quite a bit. The table shows maximum
and minimum bounds
based on linear programming in the full five-dimensional parameter space, as well as maximal reach
values. The latter are largely controlled by the numbers for the Crab nebula, for which there
is by far the best data about the emissions coming from the PeV-scale electron population.

\section{Conlusion}
\label{sec-concl}

The incorporation of recent TeV $\gamma$-ray data, collected by the Tibet AS$\gamma$ and High Altitude Water
Cherenkov Observatory (HAWC) experiments, has led to some improvements in the strength of these
astrophysical constraints on the $c^{\nu\mu}$ parameters. Even the weakest bounds are now at the
$7\times 10^{-16}$ level. There are definitely
prospects for further improvement in the future, but there are also
hard limits on how far the improvement may ultimately go.

In all probability, there are many radiating sources that have populations of PeV-energy electrons.
However, only for the relatively nearby and extensively studied Crab nebula have the synchrotron
and IC emissions attributable to these electrons been mapped out all the way up to the very highest
energies. More accurate and extended observations of other high-energy sources may continue to strengthen
these bounds; with observations of the radiation from PeV electrons in multiple well-placed sources, we
could potentially push all the coupled bounds down to the $10^{-19}$ level or better.

On the other hand, it may be that this will be the last iteration of astrophysical bounds
that remain competitive with laboratory tests of the electron $c^{\nu\mu}$---except for
the unique case of $c_{TT}$. Clock tests have improved
many orders of magnitude over the last decade, and the laboratory work is not ultimately limited
by the same factors as the astrophysical bounds. If there are no significant populations, anywhere in the
observable universe, of energetic
electrons with energies higher than a few PeV, then astrophysical bounds based on observations of
populations of radiating electrons are not going to be able to improve beyond a certain point.
However, laboratory bounds on $c_{TT}$ are notoriously difficult to establish, and astrophysical observations
will probably continue to play a central role in constraining isotropic forms of boost invariance violation.
For constraints on non-minimal SME terms, which grow in relative importance with energy, large
astrophysical energies will obviously also continue to be useful, and already the new
$\gamma$-ray telescope data is being put to work constraining isotropic, higher-dimensional Lorentz-violating
operators~\cite{ref-albert2}.

\end{document}